\documentclass[aps,prd,twocolumn,superscriptaddress,amsmath,showpacs]{revtex4-1}
\usepackage{graphicx}
\usepackage{dcolumn}
\usepackage{bm}
\usepackage{natbib}
\newcommand{\be}{\begin{equation}}
\newcommand{\ee}{\end{equation}}
\newcommand{\bea}{\begin{eqnarray}}
\newcommand{\eea}{\end{eqnarray}}

\newcommand{\Mpc}{{\rm ~Mpc}}

\urldef\mgcamb\url{http://www.sfu.ca/~aha25/MGCAMB.html}

\begin{document}
%opening
\title{New Constraints On The Dark Energy Equation of State}

\author{Najla Said}
\affiliation{Physics Department and INFN, Universit\`a di Roma ``La Sapienza'', Ple Aldo Moro 2, 00185, Rome, Italy}

\author{Carlo Baccigalupi}
\affiliation{SISSA/ISAS, Via Bonomea 265, 34136, Trieste, Italy}
\affiliation{INAF-Osservatorio Astronomico di Trieste, Via G.B. Tiepolo 11, I-34131 Trieste, Italy}
\affiliation{INFN, Sezione di Trieste, Via Valerio 2, I-34127 Trieste, Italy}

\author{Matteo Martinelli}
\affiliation{SISSA/ISAS, Via Bonomea 265, 34136, Trieste, Italy}
\affiliation{INAF-Osservatorio Astronomico di Trieste, Via G.B. Tiepolo 11, I-34131 Trieste, Italy}
\affiliation{INFN, Sezione di Trieste, Via Valerio 2, I-34127 Trieste, Italy}

\author{Alessandro Melchiorri}
\affiliation{Physics Department and INFN, Universit\`a di Roma ``La Sapienza'', Ple Aldo Moro 2, 00185, Rome, Italy}

\author{Alessandra Silvestri}
\affiliation{SISSA/ISAS, Via Bonomea 265, 34136, Trieste, Italy}
\affiliation{INAF-Osservatorio Astronomico di Trieste, Via G.B. Tiepolo 11, I-34131 Trieste, Italy}

\begin{abstract}
We combine recent measurements of Cosmic Microwave Background Anisotropies, Supernovae luminosity distances
and Baryonic Acoustic Oscillations to derive constraints on the dark energy equation of state $w$ in
the redshift range $0<z<2$, using a principal components approach. We find no significant deviations from the expectations of a cosmological constant.  However, combining the datasets we find slight indication for $w<-1$ at low redshift, thus highlighting how these datasets prefer a non-constant $w$. Nevertheless the cosmological constant is still in agreement with these observations, while we find that some classes of alternative models may be in tension with the inferred $w(z)$ behaviour.
\end{abstract}


\maketitle

\section{Introduction} \label {sec:intro}

One of the main goal of modern cosmology is to determine the nature of the 
dark energy component that is sourcing  the late time accelerated expansion of the universe.

Since its discovery from measurements of luminosity distances of type Ia Supernovae
in 1998~\cite{Perlmutter:1998np,Riess:1998cb},  cosmic acceleration has been confirmed by several 
independent cosmological data. In particular 
 measurements of the Cosmic Microwave Background and galaxy distribution, have
not only confirmed its presence but also helped in clarifying its nature.

Constraints on the dark energy equation of state $w$,
that is the ratio between dark energy pressure and density, assumed
as constant with redshift, have slightly improved over the last decade. 
For example, from the early analysis of \cite{earlyw}
where the constraint of $w< -0.6$ at $95 \%$ c.l. was reported,
the very recent analysis of \cite{wigglez} gives $w=-1.01\pm0.085$, at
$68 \%$ c.l., i.e. an improvement of a factor $2.35$ in the constraint
in about a decade of observations. These recent observations have
provided decisive evidence against dark energy models with values of
$w$ in the range $-0.8< w < -0.66$ as domain walls (see e.g. \cite{conversi})
or models based on extra dimensions (see e.g. \cite{DGP} and references therein).
Other models as "geometric" dark energy and "thawing" dark energy show, albeit
not at a statistically significant level, a better agreement with observations (see \cite{rubin}).

Up to now, data are in good agreement with the standard cosmological model, where the acceleration 
is sourced by the cosmological constant  $\Lambda$~\cite{Komatsu2011}, and therefore the equation of state is $w=-1$. However, recent experiments have reached
a sensitivity that allows to test other characteristics of the dark energy equation of
state and not only its value at very recent times. It is indeed possible to reconstruct the 
redshift dependence of $w$, a key feature to distinguish between the cosmological constant, 
which predicts a constant $w$, and alternative models which generally 
give an effecitve equation of state parameter that varies with redshift (see e.g. \cite{Yoo:2012ug,Tsujikawa:2010zza}).\\

In this paper we investigate the power of currently available cosmological data to constrain $w(z)$ at different 
redshifts, adopting the principal component analysis already applied to the equation of state of dark energy in previous studies (see e.g. \cite{huterer2004,serra2009,zunckel2010,zhao2012}).\\

The paper is organized as follows. In Section~\ref{sec:data} we list the datasets used in our analysis, while in Section~\ref{sec:method} we describe the methodology. The results
of our Markov Chain - Monte Carlo analysis are presented in Section~\ref{sec:results} while we 
draw our conclusions in Section~\ref{sec:conclusions}.

\section{Data}\label{sec:data}

We analyze a large sample of cosmological datasets.
For the supernovae SN-Ia luminosity distance we consider 
the Union1 compilation~\cite{Kowalski2008}, Union2~\cite{Amanullah2010} and SNLS~\cite{Riess2004,Astier2006}.
We then combine them separately with the CMB observations coming from seven years of observations from
the WMAP satellite \cite{Komatsu2011}.

Together with the WMAP and SNLS we consider different datasets for Baryonic Acoustic Oscillations (BAO) surveys
combined in the following way:

\begin{itemize}
  \item{run1} \\
SDSS-dr7 at z=0.20, 0.35~\cite{Percival2010} in form of $d_s(z)$, WiggleZ at z=0.44, 0.60, 0.73~\cite{Blake2011} in form of $A_s(z)$ ;
  \item{run2} \\
6dFGRS at z=0.1~\cite{Beutler2012}, WiggleZ at z=0.44, 0.60, 0.73~\cite{Blake2012} all in form $d_s(z)$;
  \item{run3} \\
WiggleZ at z=0.44, 0.60, 0.73 all in form $d_s(z)$;
  \item{run4} \\
6dFGRS at z=0.1, SDSS-dr7 at z=0.20, 0.35, WiggleZ at z=0.44, 0.60, 0.73 all in form $d_s(z)$.
\end{itemize}

In run1 we use as covariance matrix the one indicated in~\cite{Percival2010} for SDSS-dr7 points and the one shown in~\cite{Blake2011} for WiggleZ points. In the other runs we use always the covariance matrix of~\cite{Percival2010} for SDSS-dr7 where used, while for 6dFGRS and WiggleZ we use their parts of the covariance matrix from~\cite{Hinshaw2012}.
Finally we add the most recent measurements for $H_0$ from the Hubble Space Telescope~\cite{Riess2011} and the $H(z)$ dataset from Moresco et al.~\cite{Moresco2012}.

\section{Data Analysis Method} \label{sec:method}

The analysis method we adopt is based on the publicly available Monte Carlo
Markov Chain package \texttt{cosmomc} \cite{Lewis2002} with a convergence
diagnostic done through the Gelman and Rubin statistic.

We sample the following six-dimensional standard set of cosmological parameters,
adopting flat priors on them: the baryon and cold dark matter densities
$\Omega_{\rm b}$ and $\Omega_{\rm c}$, the Hubble constant $H_0$,
the reionization optical depth $\tau$, the scalar spectral index $n_S$, and the overall normalization of the
spectrum $A_S$ at $k=0.002\Mpc^{-1}$. We consider purely adiabatic initial conditions and
we impose spatial flatness.

The dark energy equation of state, as discussed above, is sampled 
in six redshift bins, $w_i(z)\,(i=1,2,..6)$,
 at six redshifts, $z_i\in{[0.0,0.25,0.50,0.85,1.25,2.0]}$
equally spaced in $ln(a)$.  Including 
more than six bins does not significantly improve the constraints.

In order to have $w(z)$ as a smooth and continuous function, we interpolate between
these values with a hyperbolic tangent function defined as:
\newline

      $\left\{
      \begin{array}{l r}
         w(z)=w_i & z=z_i\\
         w(z)= w_i + \delta_w +\delta_w tanh(\frac{\delta_z - z}{s}) & z\in [z_i,z_{i+1}]  \\
         w(z)=-1 & z\geq z_6
      \end{array}
      \right.$
     \\

where $A_i$ is an amplitude factor, $\delta_z$ is the half value of $(z_{i+1}-z_i)$, $\delta_w$ is the half value of $(w_{i+1}-w_i)$ and $s$ is a smoothing parameter.
The presence of dark energy perturbations has to be accounted for when the dark energy equation of state is constrained using cosmological probes that are sensitive to density perturbations. This is done by using a modified version of the publicly available code CAMB~\cite{Lewis:1999bs}, that evolves the dark energy perturbations also for values of $w<-1$, avoiding singularities by using the Parametrized Post-Friedmann prescription for dark energy suggested by Hu~\cite{Hu2007,Hu2008,Feng2008}.

Once the values of the parameters ${\bf w}={w_i}$ are determined, we must deal with the fact that they are correlated, which means that their covariance matrix is not diagonal. To obtain an uncorrelated estimation for their value we can make a rotation in the parameters space so to chose a different basis for the ${\bf w}$ in which their covariance matrix is diagonal. This can be done following the procedure suggested in~\cite{huterer2004,serra2009, Sarkar2007}.
Using CosmoMC we derive the covariance matrix, $\textbf{C}=(w_i-\langle w_i\rangle)(w_j-\langle
w_j\rangle)^T\equiv \langle {\bf w} {\bf w}^T\rangle-\langle {\bf
w}\rangle\langle {\bf w}^T\rangle$, and then we invert it to obtain the Fisher matrix, which can be rewritten also as 
${\bf C}^{-1} \equiv {\bf F}={\bf O^T\,\Lambda\,O}$. Here ${\bf \Lambda}$ is the diagonalized inverse covariance for the transformed bins. The vector of the uncorrelated parameters ${\bf q}$ is obtained using the rows of the tranformation matrix as weights ${\bf q}={\bf Ow}$. We can now define $\tilde{{\bf W}}$ so that $\tilde{{\bf W}^T}\tilde{{\bf W}}={\bf F}$, and, as pointed out by Hamilton~\cite{Hamilton}, there are infinitely many choices for ${\tilde{\bf W}}$. 
Following~\cite{huterer2004} we choose as weight matrix $\tilde{{\bf W}}={\bf O^T\Lambda^{\frac{1}{2}}O}$, where the rows are normalized 
to unity, and we apply it to obtain the uncorrelated parameters. 

\begin{figure}[htb!]
\begin{center}
\includegraphics[width=6cm,angle=-90]{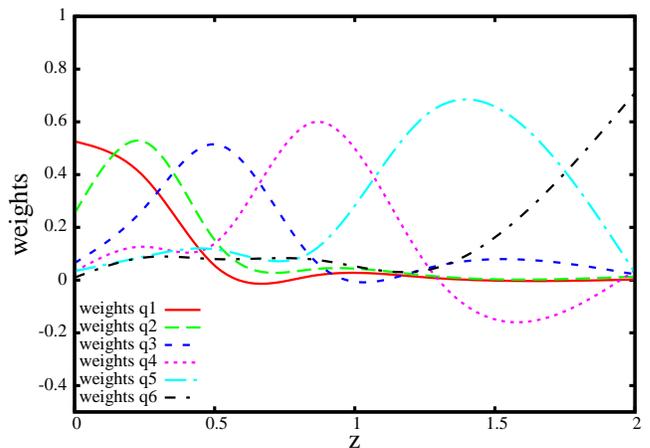}
\caption{Typical weights used to obtain the principal components $q_i$ starting from the values of the $w_i$. The weights plotted are those obtained by PCA decorrelation for run2 of Table \ref{baotable}. }  
\label{pesifigu}
\end{center}
\end{figure}

\begin{figure}[htb!]
\begin{center}
\includegraphics[width=6cm,angle=-90]{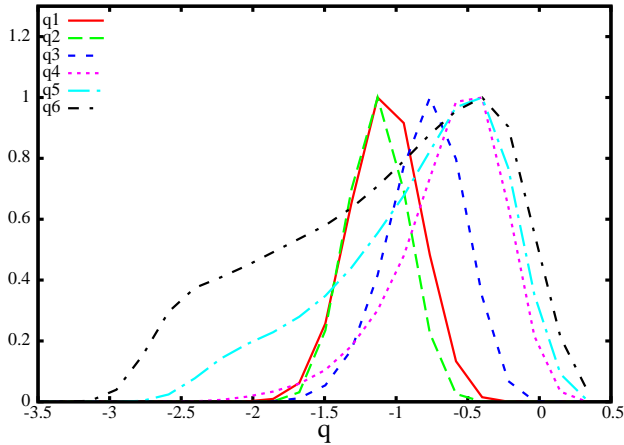}
\caption{Typical likelihoods for the principal components $q_i$. This results are obtained with PCA decorrelation for run2 of Table \ref{baotable}. }  
\label{qlikfigu}
\end{center}
\end{figure}

In other words we run CosmoMC with the ${\bf w}$ vector to obtain its covariance matrix. After that we derive ${\bf q}$, the principal components vector, by using the procedure explained before, and then run again the statistics to evaluate its marginalized values and errors. In Figure \ref{pesifigu} we show typical weights to obtain the ${\bf q}$ from a linear combination of ${\bf w}$
while in Figure \ref{qlikfigu} we show the likelihoods of the six principal components $q_i$ for a typical run. It can be noticed how, as expected, the lower redshift components are better constrained than the higher redshift ones.

\section{Results}\label{sec:results}

\subsection{WMAP+SNIa}

\begin{figure}[htb!]
\begin{center}
\includegraphics[width=6cm,angle=-90]{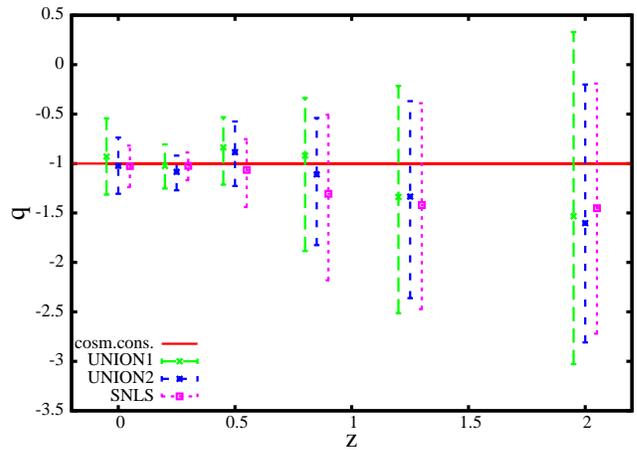}
\caption{Constraints at $95 \%$ c.l. on $q_i$ from WMAP CMB data combined with different supernovae
catalogs: Union1, Union2 and SNLS. As we can see all the datasets provide constraints on $q_i$ that
are consistent with the predictions of a cosmological constant in between two standard deviations.}  
\label{snfigu}
\end{center}
\end{figure}

We first investigate the constraints on the dark energy equation of state combining WMAP with different supernovae datasets
and we report the results of this analysis in Table \ref{sntable} and Figure \ref{snfigu}.

As we can see, all the values for the $q_i$ are consistent with the cosmological constant
case with $q_i=-1$ in between two standard deviations. However, if we look at the value of
$q_3$, around $z\sim 0.5$, the SNLS survey appears as more compatible with $w=-1$ than the Union1 and
Union2 catalogs that are on the contrary providing slightly larger values of
$q_3$. Moreover the SNLS gives stronger constraints at lower redshifts respect to
Union1 and Union2 while appears preferring a value of $q_4 < -1$ at $z\sim0.85$ whereas
Union1 and Union2 are more compatible with a cosmological constant in this redshift range.

If we look at the constraints on the other parameters in Table \ref{sntable}
we also see that the three datasets provide slightly different values
for the Hubble constant $H_0$: the SNLS survey indicates 
larger values, with $H_0\sim 72.5 km/s/Mpc$, while the Union2 and 
Union1 catalogs prefers smaller values with $H_0\sim 70.4 km/s/Mpc$ and 
$H_0\sim 68.2 km/s/Mpc$ respectively. All these values of $H_0$ are however 
well consistent in between two standard deviations.
The value of the dark energy density appears also as larger in the SNLS survey respect to the
value obtained from the Union2 and Union1 datasets, the latter providing the smallest value. 
Since we are considering a flat universe this means that SNLS data is preferring a lower matter 
density respect to Union2 and Union1.

\begin{table*}
\caption{WMAP+SNIa}
\begin{center}
\begin{tabular}{|c|c|c|c|}
\hline\hline
Parameter & Union1 & Union2 & SNLS \\
\hline
$\Omega_bh^2$ &$0.0222\pm0.0011$ &$0.0222\pm0.0011$ &$0.0222\pm0.0011$ \\
$\Omega_{\rm c}h^2$ &$0.113\pm0.011$ &$0.113\pm0.011$ &$0.113\pm0.010$\\
$H_0$ &$68.2\pm6.1$ &$70.4\pm6.1$ &$72.6\pm7.0$ \\
$n_s$ &$0.963\pm0.028$ &$0.963\pm0.026$ &$0.962\pm0.027$ \\
$log[10^{10} A_s]$ &$3.209\pm0.091$ &$3.212\pm0.085$ &$3.213\pm0.084$\\
$A_{SZ}$ &$1.0\pm1.2$ &$1.0\pm1.2$ &$0.9\pm1.1$\\
$q_6(z=2.00)$ &$>-3.0$ &$-1.6^{+1.4}_{-1.2}$ &$-1.5\pm1.3$\\
$q_5(z=1.25)$ &$-1.3^{+1.1}_{-1.2}$ &$-1.3\pm1.0$ &$-1.4^{+1.0}_{-1.1}$\\
$q_4(z=0.85)$ &$-0.92^{+0.58}_{-0.97}$ &$-1.11^{+0.57}_{-0.71}$ &$-1.31^{+0.80}_{-0.87}$\\
$q_3(z=0.50)$ &$-0.84^{+0.30}_{-0.38}$ &$-0.89^{+0.31}_{-0.34}$ &$-1.06^{+0.31}_{-0.37}$\\
$q_2(z=0.25)$ &$-1.02^{+0.22}_{-0.23}$ &$-1.08^{+0.16}_{-0.19}$ &$-1.02^{+0.14}_{-0.15}$\\
$q_1(z=0.00)$ &$-0.93^{+0.39}_{-0.38}$ &$-1.02^{+0.29}_{-0.28}$ &$-1.03\pm0.21$\\
\hline
$\Omega_{\Lambda}$ &$0.708\pm0.061$ &$0.725\pm 0.059$ &$0.742\pm0.057$\\
$t_0/Gyr$ &$13.84\pm0.32$ &$13.79\pm0.31$ &$13.72\pm0.32$\\
$\Omega_m$ &$0.292\pm0.061$ &$0.275\pm0.059$ &$0.258\pm0.057$ \\
$\theta$& $1.0380\pm0.0054$ &$1.0380\pm0.0052$ & $1.0382\pm0.0052$ \\
\hline\hline
$\chi^2$ &$7776.58$ &$7999.48$ &$7586.84$\\
\hline
 \end{tabular}
 \caption{Constraints at $95 \%$ confidence level for a WMAP analysis considering different supernovae
 datasets. The SNLS survey provides constraints that are more consistent with the $\Lambda$-CDM case.}
 \label{sntable}
 \end{center}
 \end{table*}

\subsection{WMAP+SNLS+BAO}

\begin{figure}[htb!]
\begin{center}
\includegraphics[width=6cm,angle=-90]{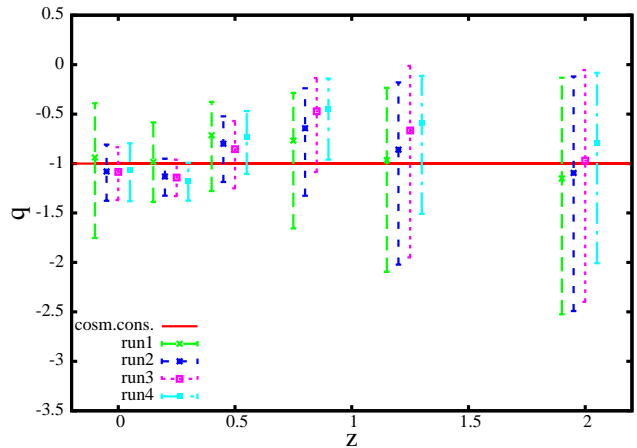}
\caption{Constraints on the dark energy equation of state from a combined analysis of the 
WMAP+SNLS dataset with different BAO datasets as described in the text. The run2 provides the
lowest chi-square value. There is a broad agreement with a cosmological constant, however values
at redshift $z\sim0.25$ are in better agreement with $w<-1$ while values at $z\sim0.5$ prefer
$w>-1$.}  
\label{baofigu}
\end{center}
\end{figure}

Since the SNLS dataset appears as the most consistent with a cosmological constant
 we perform several analysis combining the WMAP+SNLS data with four different choices for the 
 BAO datasets as specified in the previous Section.
The results on the parameters are reported in Table \ref{baotable} while the constraints
on $q_i$ are reported in Figure \ref{baofigu}.

\begin{table*}
\caption{WMAP+SNLS+BAO}
\begin{center}
\begin{tabular}{|c|c|c|c|c|}
\hline\hline
Parameter & run1 & run2 & run3 & run4 \\
\hline
$\Omega_bh^2$ &$0.0220\pm0.0011$ &$0.0221\pm0.0011$ &$0.0221\pm0.0011$ &$0.0222\pm0.0012$ \\
$\Omega_{\rm c}h^2$ &$0.119\pm0.012$ &$0.116\pm0.011$ &$0.117\pm0.012$ &$0.117\pm0.012$ \\
$H_0$ &$64.3\pm3.5$ &$65.5\pm3.8$ &$64.4\pm4.2$ &$63.6\pm3.0$ \\
$n_s$ &$0.957\pm0.029$ &$0.960\pm0.028$ &$0.960\pm0.029$ &$0.962\pm0.031$ \\
$log[10^{10} A_s]$ &$3.245\pm0.098$ &$3.229\pm0.093$ &$3.229\pm0.095$ &$3.22\pm0.10$ \\
$A_{SZ}$ &$0.9\pm1.1$ &$1.0\pm1.1$ &$1.0\pm1.1$ &$1.0\pm1.1$ \\
$q_6(z=2.00)$ &$>-2.7$ &$-1.10^{+0.98}_{-1.40}$ &$-0.97^{+0.92}_{-1.43}$ &$-0.79^{+0.71}_{-1.22}$ \\
$q_5(z=1.25)$ &$-0.75^{+0.69}_{-1.45}$ &$-0.86^{+0.68}_{-1.16}$ &$-0.67^{+0.65}_{-1.28}$ &$-0.59^{+0.47}_{-0.92}$ \\
$q_4(z=0.85)$ &$-0.52^{+0.40}_{-0.71}$ &$-0.64^{+0.40}_{-0.68}$ &$-0.47^{+0.34}_{-0.62}$  &$-0.45^{+0.31}_{-0.51}$ \\
$q_3(z=0.50)$ &$-0.84^{+0.28}_{-0.41}$ &$-0.80^{+0.28}_{-0.39}$ &$-0.85^{+0.28}_{-0.40}$ &$-0.73^{+0.26}_{-0.37}$ \\
$q_2(z=0.25)$ &$-1.21^{+0.18}_{-0.19}$ &$-1.13^{+0.18}_{-0.19}$ &$-1.14^{+0.18}_{-0.19}$ &$-1.17^{+0.18}_{-0.20}$ \\
$q_1(z=0.00)$ &$-1.06^{+0.27}_{-0.30}$ &$-1.08^{+0.27}_{-0.30}$ &$-1.08^{-0.25}_{+0.29}$ &$-1.06^{+0.27}_{-0.32}$ \\
\hline
$\Omega_{\Lambda}$ &$0.659\pm0.035$ &$0.676\pm0.040$ &$0.665\pm0.048$ &$0.656\pm0.035$ \\
$t_0/Gyr$ &$14.1\pm0.28$ &$14.01\pm0.28$ &$14.08\pm0.30$ &$14.11\pm0.28$ \\
$\Omega_m$ &$0.341\pm0.035$ &$0.324\pm0.040$ &$0.335\pm0.048$ &$0.344\pm0.035$ \\
$\theta$& $1.0370\pm0.0052$ &$1.0375\pm0.0054$ & $1.0373\pm0.0054$ &$1.0374\pm0.0055$ \\
\hline\hline
$\chi^2$ &$7595.08$ &$7591.30$ &$7593.98$ &$7597.02$ \\
\hline
 \end{tabular}
 \caption{Constraints at $95 \%$ confidence level for a WMAP+SNLS analysis considering different 
 combinations of BAO datasets (see text). BAO data clearly prefer a lower 
 Hubble constant around $H_0\sim 65 km/s/Mpc$.}

 \label{baotable}
 \end{center}
 \end{table*}

We can derive the following conclusions:

\begin{itemize}

\item There is a slight indication for $w< -1$ around the second bin $q_2$ at $z\sim 0.25$. This deviation,
already noticed in \cite{zhao2012}, is at about two standard deviations and appears in all the
BAO data combinations considered. The values of $q_3$ in the adjacent bin at $z\sim 0.50$ are on the
contrary larger than $-1$ at about one standard deviation. The agreement with a cosmological
constant is worse respect to the WMAP+SNLS case.

\item When the BAO data is included, the preferred value for the Hubble constant is significantly
smaller respect to the WMAP7+SNLS case. The values for $H_0$ are in the range $64-65 km/s/Mpc$.
These values are in strong tension with the HST result of $H_0=73.8\pm2.4 km/s/Mpc$ of \cite{Riess2011}
and may indicate the need for new physics or, more simply, the existence of systematics.\\

\item Similarly, the value of the matter density is larger when the BAO dataset is included in the
WMAP+SNLS dataset.

\item The BAO datasets are all giving consistent results. However the "run2" case provides the lowest
best fit chi-square.
\end{itemize}

\subsection{WMAP+SNLS+BAO+HST and +H(z)}

\begin{figure}[htb!]
\includegraphics[width=6cm,angle=-90]{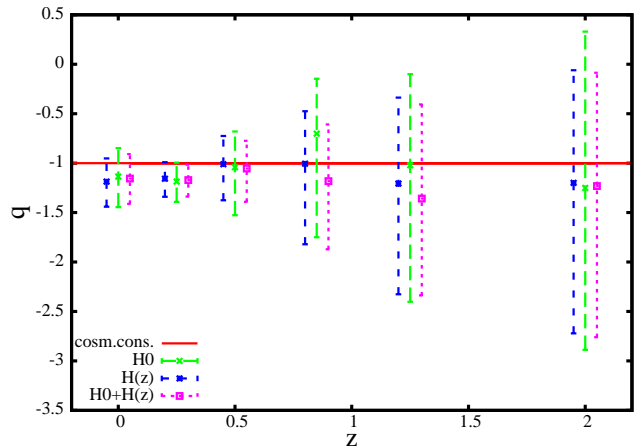}
\caption{Constraints on the dark energy equation of state from a combined analysis of the 
WMAP+SNLS+BAO dataset with the HST prior on the Hubble constant of \cite{Riess2011} 
and the determination of $H(z)$ in \cite{Moresco2012}. There is an improved agreement
with a cosmological constant respect to the WMAP+SNLS+BAO case for $q_3$ at $z\sim0.50$.}  
\label{hstfigu}
\end{figure}

Since the run2 BAO dataset provides the lowest chi-square values, we assume it, together with WMAP+SNLS, 
as our basic dataset and we now add  the HST prior on the Hubble parameter and the H(z) dataset.
The results are reported in Table \ref{hsttable} and in Figure \ref{hstfigu}.

\begin{table*}
\caption{WMAP+SNLS+BAO(run2)}
\begin{center}
\begin{tabular}{|c|c|c|c|}
\hline\hline
Parameter & +HST & +H(z) & +H(z)+HST \\
\hline
$\Omega_bh^2$ &$0.0222\pm0.0011$ &$0.0219\pm0.0011$ &$0.0221\pm0.0011$\\
$\Omega_{\rm c}h^2$ &$0.121\pm0.012$ &$0.1248\pm0.0096$ &$0.1250\pm0.0093$\\
$H_0$ &$67.2\pm3.4$ &$66.4\pm3.0$ &$68.1\pm2.7$\\
$n_s$ &$0.959\pm0.029$ &$0.950\pm0.027$ &$0.953\pm0.026$\\
$log[10^{10} A_s]$ &$3.248\pm0.097$ &$3.284\pm0.081$ &$3.277\pm0.077$\\
$A_{SZ}$ &$1.0\pm1.2$ &$0.9\pm1.1$ &$0.9\pm1.1$\\
$q_6(z=2.00)$ &$ >-2.9$ &$-1.2^{+1.1}_{-1.5}$ &$-1.2^{+1.1}_{-1.5}$\\
$q_5(z=1.25)$ &$-1.02^{+0.92}_{-1.38}$ &$-1.21^{+0.87}_{-1.12}$ &$-1.36^{+0.95}_{-0.98}$\\
$q_4(z=0.85)$ &$-0.70^{+0.56}_{-1.05}$ &$-1.01^{+0.53}_{-0.82}$ &$-1.18^{+0.57}_{-0.69}$\\
$q_3(z=0.50)$ &$-1.04^{+0.36}_{-0.48}$ &$-1.01^{+0.29}_{-0.36}$ &$-1.05^{+0.28}_{-0.34}$\\
$q_2(z=0.25)$ &$-1.19^{+0.19}_{-0.21}$ &$-1.16^{+0.17}_{-0.18}$ &$-1.18\pm0.17$\\
$q_1(z=0.00)$ &$-1.14^{+0.29}_{-0.31}$ &$-1.12^{+0.23}_{-0.25}$ &$-1.16^{+0.25}_{-0.26}$\\
\hline
$\Omega_{\Lambda}$ &$0.683\pm 0.036$ &$0.667\pm0.037$ &$0.682\pm0.033$\\
$t_0/Gyr$ &$13.92\pm0.24$ &$13.96\pm0.24$ &$13.86\pm0.22$\\
$\Omega_m$ &$0.317\pm0.036$ &$0.333\pm0.037$ &$0.318\pm0.033$\\
$\theta$& $1.0380\pm0.0052$ &$1.0368\pm0.0054$ &$1.0376\pm0.0052$\\
\hline\hline
$\chi^2$ &$7608.70$ &$7611.00$ &$7619.64$\\
\hline
 \end{tabular}
  \caption{Constraints at $95 \%$ confidence level for a WMAP+SNLS+BAO analysis considering a HST prior on the
  Hubble constant of \cite{Riess2011} and the determination of $H(z)$ in \cite{Moresco2012}}

 \label{hsttable}
 \end{center}
 \end{table*}

As we can see the effect of adding the HST prior or the $H(z)$ dataset is to 
increase the value of the Hubble parameter. However the HST prior is clearly in tension
with the WMAP+SNLS+BAO dataset as showed by the increased value of the chi-square.

Both the HST prior and the $H(z)$ dataset render
the value of $q_3$ more compatible with predictions of a cosmological constant ($w=-1$).
In general, with the exception of the value of $q_2$ at $z\sim0.25$ that prefers values
such that $w<-1$, there is a general agreement with a cosmological constant.

\subsection{Comparison with alternatives to $\Lambda$CDM}

\begin{figure}[htb!]
\begin{center}
\includegraphics[width=10cm]{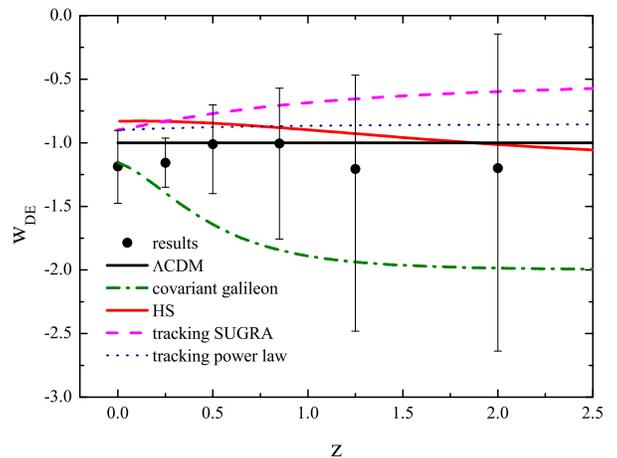}
\caption{Comparison of the constraints on the dark energy equation of state derived from a WMAP+SNLS+BAO+$H(z)$ dataset
with several dark energy models as covariant galileon, Hu-Sawicki, tracking SUGRA and tracking power-law. Most of these
models are in tension with the data mostly because of the low redshift values of $q_1$ and $q_2$ that point to
a cosmological constant or to $w(z\sim0.25) < -1$. The tracking power law model reproduces the low redshift behaviour but
is in disagreement with the constraints at higher redshifts.}  
\label{confronto}
\end{center}
\end{figure}

In order to better picture the impact of our analysis we compare the constraints on the equation of state
 obtained using WMAP+SNLS+BAO+$H(z)$ reported in the previous Section, 
  with some representative dark energy and modified gravity models alternative to $\Lambda$CDM.\\
In Fig.\ref{confronto} we show how our results on $q_i$ are in tension with the typical $w(z)$ behaviour of models such as the Hu-Sawicki $f(R)$ model (HS) \cite{Hu2007}, the covariant galileon model \cite{DeFelice:2011aa} and tracking models \cite{Brax:1999yv,Liddle:1998xm}. It is possible to notice how the tension is related to the low redshift bins; in particular, we can see how both tracking and HS models predict a $w(z)>-1$ at low redshifts colliding with our results which slightly prefer $w<-1$; the covariant galileon model instead does give $w<-1$ however, while it can fit our first two redshift bins, it has tension with values in the higher redshift bins.\\
We want to stress that in this paper we did not obtain constraints on these theories; our purpose is to show how a detailed analysis of currently available datasets can in principle rule out models that are still considered viable. In particular, while we know that the covariant galileon model is already disfavoured by other observations \cite{Nesseris:2010pc}, tracking models and HS modified gravity are still viable for some interval of the parameter space \cite{Martinelli:2009ek,Baccigalupi:2002xu}. However, they  appear to be in tension with the  combination of datasets  considered in our analysis, motivating a detailed full analysis that will be the matter of an upcoming paper.

\section{Conclusions}\label{sec:conclusions}

In this paper we have presented updated constraints on the dark energy equation of state from a large collection
of cosmological datasets. We have found that while for a WMAP+SNLS case a cosmological constant is in very good
agreement with the data, when also BAO data are considered some indications are present for a time
evolution of the dark energy equation of state. In particular the values around $z\sim0.25$ prefer
$w <-1$ above one standard deviation.

Moreover, BAO data, when analyzed in our extended parametrization of the dark energy component, prefer a low value 
for the Hubble constant with $H_0 \sim 65 km/s/Mpc$ in clear
tension with the measured HST value of \cite{Riess2011}.

Finally, we have compared our results with the theoretical predictions of dark energy and
modified gravity models that predict a time varying equation of state, finding that several 
 models do not reproduce the $w(z)$ trend obtained with our analysis. This comparison suggests 
that the combination of datasets considered in our analysis can offer powerful constraints on alternatives to the cosmological standard model. We intend to perform a full analysis of the reduction of the allowed parameters space for the latter,  when they are constrained directly, in an upcoming paper. 

\subsection*{Acknowledgements}

We would like to thank Paolo Serra for crucial help at the beginning of this project,
Martina Gerbino and Luca Pagano. AM work is supported by PRIN-INAF 2009 "Astronomical probes of Fundamental physics". 
AS is supported by a SISSA Excellence Grant. CB acknowledges partial support from the INFN-PD51 initiative.

\end{document}